\documentstyle[aps,eqsecnum,epsf]{revtex}
\begin{document}

\title{\bf Growth of three-dimensional structures by atomic 
deposition on surfaces containing defects : simulations and theory}

\author{Pablo Jensen (a), Hern\'an Larralde (b), Muriel Meunier (a)
and Alberto Pimpinelli (c)} \address{(a) D\'epartement de Physique des
Mat\'eriaux, Universit\'e Claude Bernard Lyon-1, 69622 Villeurbanne
C\'edex, FRANCE;\\ (b) Instituto de F\'{\i}sica, Lab. de Cuernavaca,
Apdo. Postal 48-3, C.P. 62251, Cuernavaca, Mor., MEXICO;\\ (c) LASMEA, 
Universit\'e Blaise Pascal,Les C\'ezeaux,F-63177 Aubi\`ere Cedex, FRANCE}

\maketitle

\begin{abstract}
We perform a comprehensive study of the formation of three dimensional
(pyramidal) structures in a large range of conditions, including the
possible {\it evaporation} of adatoms from the surface and the
presence of {\it surface defects}.  We compare our computer
simulations to theoretical calculations of the growth and find good
agreement between them. This work clarifies precedent studies of three
dimensional growth and predicts the island size distributions obtained
in the different regimes. Finally, we show how our analysis can be used
to interpret experimental data.

\end{abstract}


\newpage

How can one grow {\it useful} thin films or nanostructures from atomic
beams?  The usual and most effective way is certainly by a carefully
controlled and inspired trial and error method. Theoreticians dream
that another possibility may exist : by knowing the detailed atomic
mechanisms that govern thin film growth, one could in principle tailor
the morphologies to the desired application.  Fortunately, there are
many other justifications to the study of these atomic mechanisms :
for example the understanding in terms of atomic mechanisms of growth
experiments carried under controlled conditions, where great care is
taken to avoid complications (contamination, surface defects, etc. see
for example \cite{mo,roder,noneq,henry,mrs95,ss96}). To be able to
interpret more complex situations demands the study of models
including many atomic processes \cite{software}. The first models 
only included deposition, diffusion of the adatoms and their irreversible
aggregation to form flat islands 
\cite{lagally,tang,villain,model,boston,bales,vipi,larecherche,laszlo}.  
Ratsch et al. \cite{ratsch} improved these models by including reversible
aggregation in order to reproduce the formation of compact islands. In
this paper we get closer to new experimental situations by including
two new ingredients : the growth of {\it three-dimensional} islands
and the presence of {\it surface defects} which act as traps for the
monomers, in presence of adatom evaporation. 

Evaporation, i.e. the possibility of desorption of adatoms from the
surface, is a feature that should be observed for any system at high
enough temperatures. In this sense, it is a phenomenon that is as
general as the rest of the ingredients of recent models of film
growth, and, as we have shown in the case of two dimensional (2D)
growth \cite{evaprb}, is capable of completely changing the
quantitative behaviour of the system.  Moreover, evaporation is
present in a number of experimental situations
\cite{henry,robins,stowell}, where 3D islands are built.  We should
add that thin film growth models which include evaporation have
already been studied using a mathematical analysis of rate equations
\cite{villain,vipi,larecherche,laszlo,stowell,venables73,venables84,stoyanov}.
  Computer
simulations of such models have, to our knowledge, never been carried
out. The point is that computer simulations represent an "exact" way
of reproducing the growth, in the sense that they avoid the mean-field
approximations of rate-equations approaches
\cite{bales,japan,mrs}. We have shown previously \cite{evaprb} 
that the mean-field
equations could lead to wrong predictions in the case of 2D growth :
is this also the case in 3D growth?

Three-dimensional (3D) structures are often observed in the growth of
heteroepitaxial and non-epitaxial films. A simple explanation of the
formation of such structures, based in the bonding energies between
adsorbed atoms and adsorbed atoms and the substrate can be found in
\cite{kern}. We note that such thermodynamic arguments are not always
trustworthy, since kinetics play an essential role in determining the
growth morphologies \cite{comsa,campbell,hache}.  We will not consider
this point here and will just simulate 3D growth in a schematic way,
forcing the adatoms to build pyramidal islands as they aggregate on
the surface. A more realistic model should reproduce the geometric
structure of the islands as a result of the different relevant
energies (edge diffusion around the island, barrier for down and
up-hill diffusion, \ldots). This is beyond the scope of the present
paper where we only want to investigate the {\it consequences} of the
formation of 3D structures on the growth characteristics (mainly
saturation island density and island size distribution as a function
of the growth parameters). The possible influence of surface defects
has been stressed repeatedly \cite{kern,rr,harsdorff}. We want to
clarify the manifestations of defects on the growth and to check
simple mathematical analysis of the growth by computer simulations.

We study here the {\it first} stages of the growth, roughly until the
number of islands on the substrate saturates. The reason is that it is
in this stage that such a simplified model can be of some help to
experimentalists who want to understand the microscopic processes
present in their experiments. These fundamental microscopic processes
are most easily detected in the first stages of the growth, since in
the subsequent stages additional processes can be involved (additional
diffusion paths, interlayer transport, geometrical details of the
lattice \ldots).

The paper is organized as follows. Section I briefly presents the
model and discusses some of its approximations. Then, in section II,
we study the growth of 3D islands, first by a simple
scaling analysis in the absence of defects, then by a more rigorous
mathematical mean-field treatment, where we also include the influence
of surface defects which act as perfect traps on the surface.  In
section III, computer simulations are used to test these analytical
predictions and to calculate the island size distributions in the
different cases. After a discussion (section IV) where we compare our
analysis to precedent studies, we show in section V how
experimental results can be analyzed using these results.

\section{Presentation of the model}

In this work we will describe the properties of a still
oversimplified submonolayer thin film growth model which includes
five important physical ingredients for these systems:

{(1)} {\it Deposition}.  We will assume that atoms are deposited at
randomly-chosen positions of the surface at a flux $F$ per unit
surface per unit time.  Atoms that happen to fall on the islands that 
are formed on the surface are accomodated at their proper pyramidal 
position (see below). 

{(2)} {\it Diffusion}.  Isolated adatoms can move in a random
direction by one diameter, or one lattice spacing, which we will take
as our unit length.  We denote by $\tau$ the characteristic time
between diffusion steps and $D=1/(4 \tau)$ the diffusion coefficient
(the atomic size is taken as the length unit).

{(3)} {\it Evaporation}. Isolated adatoms can evaporate off the
surface at a constant rate. We denote by $\tau_e$ the mean lifetime of
a free adatom on the surface.  It is also useful to define 
$X_S=\sqrt{D\tau_e}$, the mean adatom diffusion length before desorption

{(4)} {\it Aggregation.} If two adatoms come to occupy neighboring
sites, they stick irreversibly and form an island. As more adatoms
are captured, the island develops as a pyramid (see below). Islands are 
assumed to be immobile and do not evaporate.
 
{(5)} {\it Defect trapping} In some parts of the paper, we introduce 
a concentration $c$
of "defects" on the surface. These defects, which are randomly distributed
on the surface, act as perfect traps for the monomers. Therefore, a monomer
which happens to occupy a defect remains there forever and serves as a
nucleation center for island growth.

In the following, we call {\it particles} or {\it adatoms} the
isolated atoms (or monomers) that are deposited on the surface, and
{\it islands} a set of connected particles (thus excluding the
monomers).

Some remarks on the assumptions of this simple model regarding its
connection to the experiments are now addressed.

(a) {\it Island morphology}---We force the islands to 
assume a pyramidal shape by immediately moving an adatom
that touches an island to the desired position (Fig. \ \ref{pyramide}). 
The pyramidal
shape is adopted because we wish to understand experiments
carried out by one of us (MM) on the system Pd/NaCl \cite{henry}
where the islands take approximately such a shape. We note that
this assumption does not affect crucially the growth : it should
not affect the growth exponents which are only determined by the fact
that islands are three-dimensional (i.e. their mass increases as
their radius to the third power, see below).

(b) {\it Island diffusion}---We neglect in this model the possibility
for dimers, trimers or larger islands to diffuse on the
substrate. Island diffusion has been observed in experiments
\cite{cludifexp} and molecular-dynamics simulations
\cite{depristo}. The effects of 2D island diffusion on the growth of thin
films {\it without evaporation} have been addressed in
Refs. \cite{boston,japan,mrs,villain,metiu,barteltdimer}.

\section{Mathematical analysis of the growth of 3D islands}

\subsection{Qualitative description} 

Before going into the details of the calculations and their
confirmation by computer simulations, we present a simple picture of
the growth mechanisms of the submonolayer structures under
consideration. We are interested mainly on two points : the time
evolution of the island concentration on the substrate and the island
concentration at saturation as a function of the growth parameters :
flux, diffusion and evaporation times and defect concentration.

The qualitative evolution of the system is essentially as follows. The
system initiates as a clean empty surface. Monomers are then deposited
at a constant rate on the surface and are allowed to diffuse and/or
desorb (evaporate). When two monomers meet, they aggregate
irreversibly to form a static island (an island is also created when a
monomer is trapped by a defect) : this is island {\it nucleation}.  As
more of these encounters occur, the island density increases with
time. Competing with this nucleation process, monomers are also
captured by islands which then become larger. At some point, islands
are so large that they quickly capture the free monomers, which
reduces the chances of two monomers meeting to nucleate a new
island. Therefore, the number of islands grows less rapidly. Moreover,
when islands become large, they start touching (coalescing). These two
effects lead to a saturation in the number of islands. Interestingly,
the saturation is attained when the surface coverage reaches a value
close to .15, independently of the parameter values. This is
equivalent to saying that saturation is reached when the mean island
radius $R$ is a fixed fraction of the island-island distance (the
coverage $\theta$ is given by $\theta \sim N R^2 \sim (R/l_{ii})^2$
where $l_{ii}$ is the mean island-island distance).  A more detailed
discussion of the evolution of the systems can be done by
distinguishing several cases according to the relative importance of
the different phenomena : diffusion, evaporation, defect
concentration. It is useful to define a typical length scale for each
of these processes: $l_{CC}=(F \tau)^{1/7}$ is a typical distance
between islands when evaporation and defect concentration are not
important (see below), $X_S=\sqrt{D\tau_e}$ is, as defined above, the
adatom diffusion length before desorption and $\ell_{def} \sim
1/c^{1/2}$ is the typical distance between defects. Now,
depending on the relative values of these three length scales, growth
will be dominated by different mechanisms which we turn on to describe
in more detail.  Note that $l_{CC}$ corresponds to the typical
island-island distance at the {\it saturation} time, i.e. it is not
the actual island-island distance (for example at the beginning of
film growth). Therefore our following qualitative discussion is only
approximated, and the more precise calculations of later sections are
necessary to justify it.

\subsubsection{"Dirty" substrates : high defect concentration} 
        
By "dirty"substrate, we mean that the island concentration is affected
by the defect concentration. We will show that this is true only if
$\ell_{CC}$ is much larger than $\ell_{def}$, i.e. the defect
concentration is high enough, even in the absence of evaporation (if
evaporation is present, it can only decrease the number of islands,
therefore increasing the relative importance of defect
concentration). The simplest case is when the defect concentration is
very high, namely if $\ell_{def}$ is much smaller than {\it both}
$X_S$ and $\ell_{CC}$. In this case, island nucleation is completely
dominated by the trapping of adatoms by defects, which leads to two
main effects : first, the island concentration reaches its saturation
value rapidly (roughly in a time $c/F$); second, this
saturation value is $N_{sat} = c$, i.e. all defects are
occupied by islands but there are no islands created elsewhere.

Note that this situation corresponds to a "low" temperature case, when
$X_S$ is large enough.  At higher temperatures, one could have $X_S
\ll \ell_{def}$ (but still $\ell_{def} \ll \ell_{CC }$ to remain in
the "dirty" substrate case). In this case, the monomer concentration
is dominated by evaporation and island nucleation still occurs on
defects. The saturation density is still equal to the defect
concentration but the kinetics is different : the time needed to
reach saturation is roughly $t_c \sim 1/(F(1+X_S^2))$.

\subsubsection{"Clean" substrates : low defect concentration} 

Here we study the cases for which the island concentration is not
affected by the presence of defects, when the substrate is
"clean" enough. This is true when $\ell_{CC}$ is much smaller than
$\ell_{def}$, irrespective of the $X_S$ value. In principle, three
cases can then arise, depending on the relative value of $X_S$ against
these two lengths.  Our calculations will show (section \ \ref{scal})
that there are only two asymptotic regimes : complete condensation and
high evaporation. In the complete condensation (CC) case, adatoms do
not evaporate from the surface and island growth proceeds mainly by
capture of the diffusing adatoms. On the contrary, in the high
evaporation limit, the number of adatoms is limited by evaporation and
the most important growth mechanism for islands is that of direct
impingement of adatoms on top of them (the contribution from the
adatoms diffusing on the surface is negligible).

The first case corresponds to $\ell_{CC } \ll X_S$, independently of
the relative ordering of $l_{def}$ and $X_S$. Then, adatoms almost
never evaporate before aggregating (after this, they are safe since
islands do not evaporate). The monomer density rapidly grows, leading to 
a rapid increase of
island density by monomer-monomer encounter on the surface. This goes
on until the islands occupy a significant fraction of the surface,
roughly 1\%. Then, islands capture rapidly the monomers, whose density
decreases. As a consequence, it becomes less probable to create more
islands, and we see that their number increases more slowly. When the
coverage reaches a value close to 15\%, coalescence will start to
decrease the number of islands. The maximum number of islands
$N_{sat}$ is thus reached for coverages around 15\%.

The second case corresponds to the opposite situation : $\ell_{CC} \gg
X_S$, with still $\ell_{CC } \ll \ell_{def}$. This happens when
$\tau_e$ is small, for example at high temperatures. In this regime,
evaporation significantly alters the growth dynamics. The main point 
is that now the monomer
density becomes roughly a {\it constant} ($F \tau_e$), since it is now mainly
determined by the balancing of deposition and evaporation. Then,
 the number of islands increases linearly with
time (the island creation rate is roughly proportional to the square
monomer concentration). We also notice that only a small fraction
(1/100) of the monomers do effectively remain on the substrate, as
shown by the low sticking coefficient value at early times (the
sticking coefficient is the ratio of particles on the substrate (the
coverage) over the the total number of particles sent on the surface
(Ft)).  This can be understood by noting that an island of radius $R$ grows by
capturing only the monomers that are deposited within its "capture
zone" (the circle of radius $R+X_S$ centered on island's center).  The
other monomers evaporate before reaching the islands.  As in the case
of complete condensation, when the islands occupy a significant
fraction of the surface, they capture rapidly the monomers. This has
two effects : the monomer density starts to decrease, and the sticking
coefficient starts to increase. Shortly after, the island density
saturates and starts to decrease because of island-island
coalescence.   

Note that one could have in principle $X_S
\leq 1$, i.e. the adatoms evaporate before diffusing. This situation,
although apparently uncommon, is not physically impossible and it also
allows us to test our predictions over a larger range of parameters.
We note that, in contrast to what is observed for two dimensional (2d) islands
\cite{evaprb}, particles deposited on top of islands significantly contribute 
to island growth. In presence of strong evaporation,
this is actually the dominant mechanism for island growth, whereas for the 2d
case this happens only in some special regimes \cite{evaprb}.

\subsubsection{Summary of our results}
\label{summa}

We present here the summary of our results in the different limits
described above. These results are derived in detail in sections \
\ref{scal} and \ref{equ}. For each regime, we give in the order the saturation
island density $N_{sat}$, the thickness at saturation $e_{sat}$
(i.e. the thickness when the island density first reaches its
saturation value), the thickness at coalescence $e_{c}$ (i.e. the
thickness when the island density starts to decrease due to
island-island coalescence), and the scaling kinetics of the mean
radius as a function of time before the saturation island density
is reached. We recall that $l_{CC}=(F \tau)^{1/7}$ and
$X_S=\sqrt{\tau_e/\tau}$. 

\vspace{.7cm}

{\bf Clean substrate}

\vspace{.5cm}

high evaporation : $X_S \ll l_{CC} \ll l_{def}$

\hspace{2cm} $N_{sat} \sim [F\tau_e(1+X_s^2)]^{2/3}$

\hspace{2cm} $e_{sat} \sim e_{c} \sim [F\tau_e(1+X_s^2)]^{-1/3}$

\hspace{2cm} $R \sim Ft$

\vspace{.5cm}

low evaporation : $l_{CC} \ll X_S \ll l_{def}$ or $l_{CC} \ll l_{def} \ll X_S$

\hspace{2cm} $N_{max}\sim\left({F\over D}\right)^{2/7}$   

\hspace{2cm} $e_{sat}\sim e_{c}\sim \left({D\over F}\right)^{1/7}$

\hspace{2cm} $R\sim(FDt^2)^{1/9} \sim t^{2/9}$  

\vspace{1cm}

{\bf Dirty substrate}

\vspace{.5cm}

high evaporation : $X_S \ll l_{def} \ll l_{CC}$

\hspace{2cm} $N_{max}\sim c$  

\hspace{2cm} $e_{sat}\sim {1\over [1+X_s^2]}$

\hspace{2cm} $e_c\sim {1\over c^{1/2}}$

\hspace{2cm} $R\sim Ft$ 

\vspace{.5cm}

low evaporation : $l_{def} \ll X_S \ll l_{CC}$ or $l_{def} \ll l_{CC} \ll X_S$

\hspace{2cm} $N_{max}\sim c$

\hspace{2cm} $e_{sat}\sim c$

\hspace{2cm} $e_c\sim {1\over c^{1/2}}$

\hspace{2cm} $R\sim \left({Ft\over c}\right)$ for $t \leq c/F$, i.e. before saturation

\hspace{2cm} $R\sim (Ft/c)^{1/3}$  between saturation and coalescence ($c/F \leq t \leq 1/Fc^{1/2}$), see section \ref{equ}.

\subsection{Scaling arguments for defect-free ("clean") surfaces} 
\label{scal}

In this section we present simple scaling arguments that allow to find
the dependence of the maximum island density $N_{max}$ as a function
of the deposition parameters (Flux F, Diffusion time $\tau$ and
Evaporation time $\tau_e$).  These arguments were originally
formulated in \cite{villain} for the special case of growth of 2D islands
by atom deposition on a high-symmetry terrace, neglecting evaporation. Here, 
the argument is extended to the case of non-negligible evaporation. 
We recall that the atomic size is taken as the length unit.

The first stage of the argument requires the determination of the
nucleation rate per unit surface and time, $1/\tau_{\rm nuc}$. A
nucleation event takes place when an adatom meets a critical
island of size ${s^*}$. We call $N_{s^*}$ the density of
critical nuclei (clusters of size $s^*$) and, following
Refs. \cite{venables84,stoyanov}, we assume that $N_{s^*}$ satisfies
Walton's relation $N_{s^*}\sim \rho^{s^*}$. Thus,
\begin{equation}
\label{nuc1}
\frac{1}{\tau_{\rm nuc}} \approx (F+D\rho)N_{s^*}
\end{equation}

where $D=1/(4 \tau)$ is the
adatom diffusion constant. The term $FN_{s^*}$ accounts for direct
impingement of atoms onto critical islands and the second term
for critical island growth by monomer diffusive attachement.

Another, independent equation can be written down to relate the
nucleation rate and the stable ($s < s^*$) island density $N$. It 
states that in the area
$\ell_s^2 = 1/N$ occupied by an island, only one (on average)
nucleation event takes place, {\it during the time $t_c$ needed for
the growing islands to come into contact}. Thus,
\begin{equation}
\label{nuc2}
\frac{1}{\tau_{\rm nuc}} \approx \frac{N}{t_c} \;.
\end{equation}

The time $t_c$ is readily computed by knowing the growth velocity of
an island, which in turn requires the knowledge of the adatom
density. We consider in the following three situations of interest for
this paper.

\subsubsection{Negligible evaporation}.

The adatom density results in this case from a balance between 
deposition at a rate
$F$ and capture by the stable islands at a rate $D\rho N$, so that 
\begin{equation}
\label{n1}
\rho\approx  F/(DN) \approx F\ell_s^2/D \;.
\end{equation} 

The rate of growth of the volume of an island of linear size $R$ is
diffusion-limited in this case, so that $d(R^3)/dt \approx D\rho$ and
$R^3\approx Ft/N$. At $t=t_c$, $R\approx\ell_s$, and thus 
\begin{equation}
\label{tc5}
t_c\approx N\ell_s^3/F\approx 1/(FN^{1/2})\;. 
\end{equation}
Using (\ref{nuc1}) and (\ref{n1}), one finds
\begin{equation}
\label{rel4}
\frac 1{\tau_{\rm nuc}} \approx D[F/(DN)]^{s^*+1}
\end{equation}
(here, the direct impingement term is negligible).

From (\ref{nuc2}) and (\ref{tc5}),
\begin{equation}
\label{nucc7}
\frac 1{\tau_{\rm nuc}} \approx FN^{3/2}\;.
\end{equation}
Finally, (\ref{rel4}) and (\ref{nucc7}) yield \cite{venables84,stoyanov}
\begin{equation}
\label{3d1}
N\approx \left( \frac FD\right)^{2s^*/(2s^*+5)}\;.
\end{equation}

\subsubsection{Strong evaporation}.
\label{strong}

Strong evaporation means the adatoms are more likely to disappear due to
desorption than to be
captured by an island. In other words, the adatom diffusion length before
desorption,
$X_S=\sqrt{D\tau_e}$, is shorter than the average island-island distance,
$\ell_s$. In this case, the adatom density results from a balance between
deposition and desorption at a rate
$\rho/\tau_e$, so that 
\begin{equation}
\label{n2}
\rho\approx F\tau_e\approx FX_S^2/D \;.
\end{equation} 

A 3-D island grows by two mechanisms in the case of strong evaporation: (i) by
capture of the adatoms falling on the surface at a distance smaller than $X_S$;
(ii) by direct capture of all adatoms falling on it. Thus, 
${\rm d} R^3/{\rm d} t \approx
F(X_SR +X_S^2 + R^2)\approx F(R^2+X_SR)$. Solving for $R$ with the condition $R=0$ at $t=0$, 
one gets $R-X_S\ln(1+R/X_S)=Ft$. At $t=t_c$, $R\approx\ell_s>X_S$, which means 
that direct capture always dominates. It follows $Ft_c\approx\ell_s $, or 
\begin{equation}
\label{tc6}
t_c\approx 1/[FN^{1/2}]\;. 
\end{equation}
Using (\ref{nuc1}) and (\ref{n2}), one finds 
\begin{equation}
\label{rel5}
\frac 1{\tau_{\rm nuc}} \approx (F+D\rho)\rho^{s^*}\approx
F(1+X_S^2)\rho^{s^*}\approx F(1+X_S^2)(F\tau_e)^{s^*}\;.
\end{equation}

From (\ref{nuc2}) and (\ref{tc6}),
\begin{equation}
\label{nucc8}
\frac 1{\tau_{\rm nuc}} \approx FN^{3/2}\;.
\end{equation}
Finally, (\ref{rel5}) and (\ref{nucc8}) yield
\begin{equation}
\label{nmaxstrev}
N\approx \left(F\tau_e\right)^{2s^*/3}\left(1+X_S^2\right)^{2/3}\;.
\end{equation}

We discuss the preceding results in section \ \ref{discussion}.

\subsubsection{Crossover between the two preceding regimes}

We will show in this section that it is possible to derive
interpolation formulae describing the crossovers between the
no-evaporation and the strong-evaporation regimes.

To do this, we will use Burton, Cabrera and Frank's theory of step flow
\cite{papbcf}. The adatom density will be computed on a terrace bounded by two
parallel steps, at a distance $\ell$. They may be the steps in the orderd array
of a vicinal surface; in this case, we will let $\ell=d$. Or they may represent
the edges of a big enough island; in the latter case we will let
$\ell=\ell_s=1/\sqrt{N}$. 

The adatom density obeys the equation
\begin{equation}
\label{bcf}
\dot n = F+D\nabla^2n-\frac n\tau \;.
\end{equation}

The time $\tau$ is the average lifetime of an adatom. Since adatoms disappear
either by capture by islands, or by desorption, we will alternatively let
$1/\tau = DN  = D/{\ell_s^2} $ or $1/\tau= 1/\tau_e=D/{X_S^2} $. In both cases,
the notation $\kappa^2=1/(D\tau)$ will be used.

In the quasi-stationary approximation \cite{papbcf}, $\dot n \approx
0$, and choosing the origin $x=0$ at the terrace centre, equation
(\ref{bcf}) can be solved with the conditions $n(\pm \ell/2)=0$ at
boundary steps. The solution reads
\begin{equation}
\label{sol1}
n(x) = F\tau\left[ 1-\frac{\cosh(\kappa x)}{\cosh(\kappa \ell/2)}\right] \;.
\end{equation}

This formula will be needed to compute the nucleation rate (\ref{nuc1}). The
latter is an average quantity, independent of $x$. We let thus $x=0$ in
(\ref{sol1}), since the higher nucleation probability is at the terrace centre,
given the symmetry of our problem. One finds
\begin{equation}
\label{sol2}
n = F\tau\left[ 1-\frac{1}{\cosh(\kappa \ell/2)}\right] =2F\tau
\frac{\sinh^2(\kappa \ell/4)}{\cosh(\kappa \ell/2)}\;,
\end{equation}
where we used the identity $\cosh(x)-1=2\sinh^2(x/2)$.

The next task is the determination of the island density. To this end,
it suffices to consider the total island density $N$. Its time
variation is simple: $N$ increases each time a new island is
nucleated, so $(\dot N)_1 = D\rho N_{s^*}$.  On the other hand, $N$
decreases when two islands touch and coalesce.  Following the
authorities \cite{venables84,stoyanov}, we write $(\dot N)_2 = -({\rm d}
{\cal A}/{\rm d} t) N^2$, where ${\cal A}\approx R^2$ is the average
area of an island of linear size $R$.  This means that coalescence
results from binary encounters of immobile islands, whose area
increases at a rate ${\rm d} {\cal A}/{\rm d} t$. Collecting $(\dot
N)_1$ and $(\dot N)_2$ yields
\begin{equation}
\label{isl1}
 \dot N = D\rho N_{s^*}-\frac{{\rm d} {\cal A}}{{\rm d} t} N^2 \;.
\end{equation}

At stationarity, which is what we are interested in, $\dot N=0$. Note
that, by definition, coalescence takes place when ${\cal A}\approx
\ell_s^2$. One can thus write
\begin{equation}
\label{isl2}
  D\rho N_{s^*}\approx \left(\frac{{\rm d} {\cal A}}{{\rm d} t}\right)_{{\cal
A}\approx\ell_s^2} N^2
\;.
\end{equation}

The final task concerns therefore the evaluation of the growth rate of
an island.  This can be done by noting that the mass ${\cal M}$ of an
island increases either by direct capture of atoms from the beam, or
by diffusion of adatoms on the surface. Since the surface diffusion
current of adatoms is $ -D\nabla n$, assuming circular ($d=2$) or
hemispherical ($d=3$) islands of radius $R$ one can write
\begin{equation}
\label{mass}
\frac{{\rm d} {\cal M}}{{\rm d} t} \approx FR^{d-1} - D\left(\frac{{\rm d} n}{{\rm d} r}\right)_R
\;.
\end{equation}

The result can be found in a number of papers
\cite{evaprb,venables84,stoyanov,papbcf}. It makes use of Bessel
functions, which are not easy to manipulate. Approximate results,
which have the merit of reproducing the correct limiting regimes (see
below), and of allowing analytical expressions to be written down,
will be used.

The adatom density is still given by Eq.(\ref{sol2}), and the
nucleation rate by Eq.(\ref{nuc1}). The growth rate of the (projected)
area of a 3-D island has two contributions, as for 2-D islands: a
diffusion-limited contribution, which is given by
\begin{equation}
\label{gr1}
\frac {{\rm d} {\cal A}}{{\rm d} t}\vert_{R=\ell_s} \approx
FX_S\ell_s\tanh(\kappa\ell_s/2)
\;;
\end{equation}
and a direct capture contribution, 
\begin{equation}
\label{gr2}
\frac {{\rm d} {\cal A}}{{\rm d} t}\vert_{R=\ell_s} \approx F\ell_s
\;.
\end{equation}

A useful interpolation formula between (\ref{gr1}) and (\ref{gr2}) is
\begin{equation}
\label{da3}
\frac {{\rm d} {\cal A}}{{\rm d} t}\vert_{R=\ell_s} \approx
F\ell_s\left(1+\frac{X_S}{\ell_s}\right)\tanh(\kappa\ell_s/2)
\;.
\end{equation}

Finally,  (\ref{nuc2}) and (\ref{da3}) yield
\begin{equation}
\label{sc3}
\frac{\left(1+  X_S\right)^2}{\tau_e}
\left[F\tau_e\frac{\sinh^2(\kappa
\ell_s/4)}{\cosh(\kappa\ell_s/2)}\right]^{s^*+1}\approx
F\ell_s\left(1+\frac{X_S}{\ell_s}\right)\tanh(\kappa\ell_s/2) N^2
\;,
\end{equation}
or,
\begin{equation}
\label{sche13}
{\left(1+  X_S\right)^2}(F\tau_e)^{s^*}
\approx \left[\frac{\cosh(\kappa N^{-1/2}/2)}{\sinh^2(\kappa
 N^{-1/2}/4)}\right]^{s^*+1}
\tanh(\kappa N^{-1/2}/2)(1+X_SN^{1/2}) N^{3/2} \;.
\end{equation}

Multiplying both sides by $X_S^3$ yields
\begin{equation}
\label{sche14}
X_S^{2s^*+3}(1+X_S^2)(F/D)^{s^*} \approx f(X_SN^{1/2})\; ,
\end{equation}
where
\begin{equation}
f(x) =
\left[\frac{\cosh\left(\frac{1}{2x}\right)}{\sinh^2\left(\frac{1}{4x}
\right)}\right]^{s^*+1}
\tanh\left(\frac{1}{2x}\right)(1+x) x^{3}\;.
\end{equation}

Letting $N_{\rm 3Devap}=\left(FX_S^2/D\right)^{2s^*/3}\left(1+X_S^2\right)^{2/3}$
, formula (\ref{sche14}) can be cast in the form
\begin{equation}
\label{sche15}
\tilde f_{\rm 3Devap}(X_S^2N)=  X_S^2N_{\rm 3Devap} 
\;,
\end{equation}
where $\tilde f_{\rm 3Devap}(x)=[f(x^{1/2})]^{2/3}$. Inverting $\tilde f(x)$ yields the
crossover scaling function
\begin{equation}
X_S^2N=g_{\rm 3Devap}(X_S^2N_{\rm 3Devap}) \;.
\end{equation}

The function $g_{\rm 3Devap}(x)$ has the following properties:
\begin{equation}
\label{eqcross}
g_{\rm 3Devap}(x)\sim\cases{{x^{3/(5+2s^*)}} & {\rm for} ~$x\to\infty$,\cr\cr
                   {{x}} & {\rm for}~ $x\to 0$.\cr}
\end{equation}


\subsection{Rate equations} 
\label{equ}


In this section we study the growth in presence of defects using
rate equations \cite{zinsmeister}. 

We will consider that the islands are semi-spherical
``droplets'' as a result of ``fast'' large scale reacomodation of the monomers
forming the island. In addition to evaporation, we will also consider
the effects due to the presence of point defects on the surface. We
will assume that these defects serve as perfect nucleation points, and
so, as the concentration of defects increases, the system passes from
homogeneous to heterogeneous nucleation.

We can write the evolution of the density $\rho$ of monomers on the 
surface as:
$$
{d\rho\over dt}=F-\rho/\tau_e -F\rho - \sigma_o(2\rho+c_{def}) -\sigma_i N.
\eqno(1)
$$

The first term on the RHS denotes the flux of monomers onto the
surface.  The second term represents the effect of evaporation. The
third term is due to the possibility of losing monomers by effect of
direct impingement of a deposited monomers right beside a monomer
still on the surface to form an island.  The next terms represent the
loss of monomers by aggregation with other monomers, nucleation on
defects and aggregation with islands respectively. The factors
$\sigma_o$ and $\sigma_i$ are the ``cross sections'' for encounters
and are detailed below.

The number $N$ of islands will be given by:
$$
{dN\over dt}=F(\rho+c_def)+\sigma_o(\rho+c_{def})
\eqno(2)
$$

where the first term represents the formation of islands due to direct
impingement of deposited monomers onto other monomers or defects, and
the second term accounts for the formation of islands by the encounter
of two monomers diffusing on the surface, or a monomer on a defect. It
should be noted that since this description yields scaling results,
some numerical factors are omitted.

The concentration of ``free defects'' varies as
$$
{d c_{def}\over dt}=-(\sigma_o + F) c_{def}.
$$

The total island mass density ($M$) changes as
$$
{dM \over dt}=2\left[F\rho+\sigma_o\rho\right]
                  +\sigma_i N + \sigma_o c_{def} + FR^2N, \eqno(5)
$$

where the direct impingement term is proportional to the area of the
islands.

The typical island radius $R$ will be given by
$$
R\sim \left({M\over N}\right)^{1/3}, \eqno(6)
$$
and the actual island coverage will then be
$$
\Omega\sim N R^2 = (N M^2)^{1/3}. \eqno(7)
$$

The expressions for the diffusive cross sections in the presence of
evaporation were calculated in \cite{evaprb}.

$$
\sigma_i\sim\cases{{D\rho R\over X_S} & for $R>>X_S$,\cr\cr
                   {D\rho} & for $R<<X_S$.\cr}\eqno(12)
$$

Where $X_S=\sqrt{D\tau_e}$ is the typical distance traveled by a diffusing 
particle on the surface before it desorbs. And:

$$
\sigma_o\sim D\rho.\eqno(13)
$$

These mean-field cross sections only depend on the radius of the
sphere touching the surface, and not on the height of the island,
therefore they are the same for 2d or 3d particles of same radius.

As the most we can expect to obtain from this description are the
scaling behaviors, we focus only on the extreme limiting cases of the
system described by the rate equations.

For the case of negligible evaporation and high enough initial
concentration of defects, we expect the equations to reduce to 

$$
{d\rho\over dt}\sim F - D\rho( c_{def}+ N).  
$$ 

$$ 
{d N\over dt} \sim D\rho (\rho+c_{def}) 
$$ 
$$ 
{d c_{def}\over dt}=-D\rho c_{def}.  
$$
Adding the second and last equations, we have: 
$$ 
{d (N+c_{def})\over dt} \sim D\rho^2.  
$$ 
For high enough initial concentration of defects $c$, 
we can write 

$$
N(t)\sim c-c_{def}(t)+D({F\over Dc})^2t\sim c
$$
$$
\rho\sim {F\over Dc}
$$
$$
c_{def}(t)\sim ce^{-{Ft\over c}}.
$$

The island density at saturation is therefore equal to the initial
defect concentration $c$. To find what is "high enough" for the defect
concentration, we note that nucleation events start to reduce the
number of islands roughly when the island size is of the order of the
distance between islands, or equivalently, when the coverage reaches a
constant value. Since the deposited mass grows as $Ft$ (evaporation is
negligible), the surface coverage will be given by

$$
\Omega\sim N^{1/3}(Ft)^{2/3}\sim (cF^2 t^2)^{1/3}
$$

Thus, the coalescence time is $t_c\sim 1/(Fc^{1/2})$ and 
the cross over to the ``clean system'' occurs when

$$
({F\over Dc^{5/2}})\sim c\qquad {\rm i.e.}\qquad 
c\sim (F/D)^{2/7}.
$$

It should be stressed that in this defect dominated, or "dirty",
regime the island number density saturates far before coalescence,
indeed the saturation time $t_s$ can be estimated by the characteristic
time for defect ocupation: $t_s\sim c/F$. After $t_s$, the island
density remains essentially constant and the islands grow (the typical
island radius can be easily shown to grow as $R\sim (Ft/c)^{1/3}$)
until coalescence.

At high evaporation rates the aggregation of mass on the surface is
dominated by ``direct impingement'' of particles on the islands, and
the concentration of monomers on the surface is determined by the
balance between deposition and evaporation. Under these conditions, at
high initial defect concentration we will have:

$$
\rho\sim F\tau_e
$$
$$
{d N\over dt}\sim F\tau_eDc_{def} + F^2\tau_e + DF^2\tau_e^2
$$
$$
{d c_{def}(t)\over dt}\sim -F(D\tau_e+1)c_{def}
$$

These equations can be solved immediately, from which we get:

$$
c_{def}\sim ce^{-[X_s^2+1]Ft]}
$$
and
$$
N\sim c[1- e^{-[X_s^2+1]Ft]}] + F^2\tau_e[X_s^2+1] t,
$$
where we have substituted $X_s=\sqrt{D\tau_e}$. 

From these expressions it is evident that the time at which N saturates
is $t_s\sim 1/(F[1+X_s^2])$, after which time, the island density
reaches the value $N\sim c$ (this, of course, assuming that the
homogeneous contributions are negligible throughout the evolution of
the system). It is then straight forward to find that the mass
deposited on the system grows as $M\sim cF^3t^3$ and the typical
island radius $R\sim Ft$.  The coverage increases as $\Omega\sim
cF^2t^2$, from which we can estimate the coalescence time
$t_c\sim1/Fc^{1/2}$. While it can be imagined that these times could
occur in the wrong order, we have checked that for this to be the case, an
initial concentration of defects larger than one would be required.

By comparing the maximum value of the subdominant homogeneous term
with $c$, we can determine that this high evaporation dirty regime
is attained when $c>>(F\tau_e[1+X_s^2])^{2/3}$.

Finally, for completeness, we sketch the derivation of the results for
the ``clean'' substrate with negligible evaporation. Under these
circumstances, the monomer density on the surface is determined by the
balance between deposition and diffusive capture by the islands on the
surface. Thus, the monomer and island densities are given by

$$
\rho \sim {F\over DN}\qquad \qquad {\rm and} \qquad \qquad 
{dN\over dt}\sim D\rho^2
\sim {F^2\over DN^2}.
$$

From these equations we find that the island density is $N\sim
(F^2t/D)^{1/3}$.  Also, as we are considering the case of negligible
evaporation, the mass deposited on the surface will be $M\sim
Ft$. From these quantities we can infer the behavior of the typical
island radius to be given by $R\sim (M/N)^{1/3}\sim (DFt^2)^{1/9}$, and
the saturation and coalescence times to be $t_c\sim t_s \sim (D/F^8)^{1/7}$,
at which times the maximum value of N is $N_{max}\sim (F/D)^{2/7}$, as
obtained in the previous section.


\section{Computer simulations}


In the following paragraphs, we test the assumptions and predictions of
the analysis given in the preceding sections. We also show results that
are not attainable from this mean-field calculations, namely the
island size distributions.

Our computer simulations generate sub-monolayer structures using the
four processes included in our model (see the introduction). Here we
take $\tau=1$ as the time scale of our problem. The monomer diffusion
coefficient is then given by $D=1/4$. We use triangular lattices (six
directions for diffusion) of sizes up to $2000 \times 2000$ with
periodic boundary conditions to limit finite size effects.

The program actually consists of a repeated loop. At each loop, we
calculate two quantities $p_{drop} = F / (F + \rho ({1\over \tau_e} +
{1\over \tau}))$ and $p_{dif} = (\rho /\tau) / (F + \rho ({1\over
\tau_e} + {1\over \tau}))$ that give the respective probabilities of
the three different processes which could happen : depositing a
particle (deposition), moving a particle (diffusion) or removing a
particle from the surface (evaporation). More precisely, at each loop
we throw a random number p ($0 < p < 1$) and compare it to $p_{drop}$
and $p_{dif}$.  If $p < p_{drop}$, we deposit a particle; if $p >
p_{drop} + p_{dif}$, we remove a monomer, otherwise we just move a
randomly chosen monomer.  After each of these possibilities, we check
whether an aggregation has taken place and go to the next loop (for
more details, see \cite{boston}).

\subsubsection{Checking the crossover scaling}

Before looking in detail into the different regimes predicted in section 
\ref{equ}, we summarize our simulation results in Fig. \
\ref{crossover}. We show there {\it all} our data for $N_{max}$ as a 
function of the parameters. Our scaling
analysis predicts that the data should fall into a single curve, given
by Equation \ \ref{eqcross}. We see that the data remarkably confirms
our analysis, over more than 30 orders of magnitude. This gives us
confidence on our entire approach and its predicted exponents, which
we now turn on to check in more detail.

We now check that the results summarized in section \ref{summa} are correct. 

\subsubsection{
Scaling of the maximum island density as a function of incident flux}

Figure \ \ref{nmaxfnoscal} shows the evolution of the maximum island
density as a function of the flux for different evaporation times.
Each of these curves is different from the others, since they
correspond to different evaporation times. However, according to our
preceding analysis, they should all present a transition from the low
evaporation regime to the high evaporation regime.  This can be
detected by a change of slope, from $N_{max} \sim F^{2/3}$ in the high
evaporation regime (solid line) to $N_{max} \sim F^{2/7}$ in the low one 
(dashed line).  Of course, this regime change does not occur for all
the curves at the same value of the Flux, since the parameter that
determines that change is not the Flux but rather $X_S^2 =
\tau_e/\tau$.  Figure \ \ref{nmaxfnoscal} shows that the results of
section \ref{summa} accurately describes the behaviour of our model, at
least concerning the Flux evolution of the maximum island density. We
now turn to the other variable, the evaporation time.

\subsubsection{Maximum island density as a function of evaporation time}

We show in Figure \ \ref{nmaxtenoscal} the dependence of the
maximum island density on $\tau_e$ ($\tau=1$). We notice that for high enough
evaporation times, the island density tends to become roughly
constant, as predicted by our calculations.  For lower values of
$\tau_e$, $N_{max}$ changes rapidly. We predict (section \ref{summa}) two
regimes : for $1 \ll \tau_e \ll F^{-1/3}$, we expect 
$N_{max} \sim \tau_e^{4/3}$,
while for $\tau_e \ll 1$, we expect $N_{max} \sim {\tau_e}^{2/3}$. This last
regime is clearly seen for the curves obtained for fluxes 
$F=10^{-6}$ and $F=10^{-4}$ (squares and diamonds respectively, the
slope 2/3 is given by the solid line).
The first regime is difficult to see for two reasons. First, we need
$X_S \gg 1$ {\it and} strong evaporation, i.e. $\ell \gg X_S$. This
means a very low island density, meaning very long computing times
and large lattices. Second, the crossovers with
the two other regimes (exponents 2/3 and 0) tend to mask the exponent
4/3. Taking a lower value for the flux 
($F=10^{-8}$, filled circles), we can see that the slope in this 
intermediate regime is larger than 2/3.

\subsubsection{Mean island radius versus time for clean substrates}
\label{sectionrt}

Our treatment predicts two limiting regimes for the power-law ($r \sim t^\beta$)
evolution of the mean island radius as a function of time : $\beta=1$ in cases
of strong evaporation and $\beta=2/9$ for complete condensation (we only treat
here the case of "clean" substrates"). Fig. \ \ref{rt}a shows that 
we observe indeed an exponent very close to 2/9=0.22 when evaporation
is negligible ($X_S=10^9, l_{CC}=37$, $X_S \gg l_{CC}$), while the exponent is 
close to 1 in the opposite case ($X_S=1, l_{CC}=14$, $X_S \ll l_{CC}$), 
see Fig. \ \ref{rt}b. Of course, intermediate 
cases can arise in experiments and the exponent is between these two
extreme values, with values around 0.5-0.6 as shown in Fig. \ \ref{rt}c 
($X_S=10, l_{CC}=27$). Note that we have defined
here the radius as $(M/N)^{1/3}$ where M is the total mass present on the
substrate and N the island density, but we have checked that similar
exponents are measured if one defines the mean radius as $\Omega/N^{1/2}$
where $\Omega$ is the surface coverage.

\subsubsection{Dynamical evolution of island density}

Here we investigate how the different microscopic mechanisms can
affect the growth kinetics. This can be an important help for
experimentalists seeking information on which processes are
actually present in their experiments \cite{revmod}. Fig. \
\ref{cinetns} confirms our analytical analysis and
shows that evaporation or the presence of surface defects can significantly
alter the time evolution of island density. If defects are present,
monomers will be trapped by them at the very beginning of the
growth and the number of islands equates rapidly the number of defects,
whatever the diffusivity of the atoms. If evaporation is present, the opposite
effect is observed : since many atoms do not contribute to the growth (they
evaporate before reaching an island), the saturation is reached for very
high thicknesses (typically $e_{sat} \gg 1 ML$).

\subsubsection{Island size distributions}

Island size distributions have proven very useful as a tool for
experimentalists to distinguish between different growth mechanisms
\cite{ss,stroscio}. By {\it size} of an island, we mean the
surface it occupies on the substrate. For the "three dimensional" particles
studied here, their projected surface is the easiest
quantity to measure by microscopy.  Note that the projected surface 
for a given mass depends on the precise shape of the islands, which is 
assumed here to be pyramidal (close to a
half-sphere). Size distributions are normalized by the mean island
size in the following way : one defines $p(s/s_m) = n_s / N_t$ as the
probability that a randomly chosen island has a surface $s$ when the
average surface per island is $s_m = \theta/N_t$, where $n_s$ stands
for the number of islands of surface $s$, $N_t$ is the total number of
islands and $\theta$ for the coverage of the surface.  It has been
shown \cite{model} that by normalizing the probabilities and plotting
$s_m * p(s/s_m)$ against $s/s_m$, one obtains a "universal" size
distribution independent of the coverage, the flux or the
substrate temperature for a large
range of their values. These size distributions can be obtained from
the simulations \cite{evaprb,japan,mrs,smilauer}.

Fig \ \ref{distrib3d} shows the evolution of the {\it rescaled} island
size distributions for three dimensional islands (pyramids) in
presence of evaporation. We recall that size means here the projected
surface of the island, a quantity which can be measured easily by
electronic microscopy. We note the same trends as for 2d islands \cite{evaprb}. 
It is clear that the distributions are significantly
affected by the evaporation, smaller islands becoming more numerous when
evaporation increases. This trend can be qualitatively understood by noting
that new islands are created continuously when evaporation is present, while
nucleation rapidly becomes negligible in the complete condensation regime. The
reason is that islands are created (spatially) homogeneously in the last case,
because the positions of the islands are correlated (through monomer diffusion),
leaving virtually no room for further nucleation once a small portion of the
surface is covered ($\theta \sim 0.05$). In the limit of strong evaporation,
islands are nucleated randomly on the surface, the fluctuations 
leaving large regions of the surface uncovered. These large regions can
host new islands even for relatively large coverages, which explains that there
is a large proportion of small ($s < s_m$) islands in this regime.

Fig \ \ref{distribdef} shows the evolution of the {\it rescaled}
island size distributions for pyramidal islands nucleating {\it on
defects}. Two main differences can be noted. First, the histograms are
significantly narrower than in the preceding case, as had already been
noted in experimental studies \cite{harsdorff}.  This can be
understood by noting that all islands are nucleated at almost the same
time (at the very beginning of growth).  The second point is that the
size distributions are sensitive to the actual coverage of the
substrate, in contrast with previous cases.  In other words, there is
no perfect rescaling of the data obtained at different coverages, even
if rescaling for different fluxes or diffusion times has been checked.

\section{Discussion}
\label{discussion}

Other authors have analyzed similar mean-field rate equations to find
the growth dynamics and maximum island density in the presence of
evaporation \cite{stowell,venables73,venables84,stoyanov}. They have
also obtained different regimes identified roughly in the same way as
in the present work. We think that there is some ambiguity in their 
definition of the different regimes. A regime - meaning a single
relation between $N_{max}$ and the deposition parameters - should be
defined only by the values of F, $\tau_e$ and $\tau$, as in Section \
\ref{summa}. This is what is required from the experimental side : given
some values of the parameters, what will happen on the surface?
Instead, previous works have introduced other parameters, such as the 
coalescence coverage, the capture cross sections, or even the island density
itself, in the characterization of the regimes. 

Besides this general remark,
we note that there is some disagreement about the
different regimes between the various authors. Stoyanov and Kashchiev 
\cite{stoyanov} find two regimes which correspond to our complete
condensation and strong evaporation (with $X_S \gg 1$) cases. We have
added here the case  $X_S \ll 1$. Venables et
al. \cite{venables84} find three different regimes. The two extreme
regimes coincide with Stoyanov's and ours, but their intermediate
regime (which predicts $N_{max} \sim (F\tau_e)^{2/5}$ is not observed in our 
simulations : this is particularly clear
in Fig. \ \ref{nmaxtenoscal} which shows no $\tau_e^{2/5}$
dependence : instead, the intermediate regime shows an exponent greater than
2/3, as predicted by Eq. \ \ref{nmaxstrev}). Actually, Venables's
intermediate regime seems physically strange because there is no 
dependence on the diffusion coefficient of the adatoms.

The time evolution of the mean radius has also been studied previously
(for a recent review, see \cite{henryrt} and references therein). 
Here, we have shown that the exponent of the radius 
versus time power-law is 1 when evaporation is important (islands grow only by
direct impingement), 1/3 when condensation is almost complete {\it and}
the number of islands is constant (for example when their concentration 
reaches the defect
concentration for "dirty" substrates) or 2/9 for clean substrates and
complete condensation (here the number of islands is never really constant
since nucleation does not stop until coalescence). 

In previous studies \cite{henryrt}, it has been predicted that
the radius shows a power-law dependence with an exponent 1/3 for 
complete condensation growth and 1 in the case of strong evaporation.
Intermediate values were found thanks to the numerical resolution of
mean-field equations of island growth \cite{henryrt}.

The experimental values are rather in the range 0.21 - 0.30 for the
Pd/MgO(100) system \cite{henryrt} and around 0.3 in other systems.
While the intermediate exponents (between 1/3 and 1) can be explained
as pertaining to the intermediate regime (see section \ref{sectionrt}), 
the values lower than 1/3 are more difficult to explain. In our analysis,
the exponents close to 2/9=0.22 can simply be explained 
(in the complete condensation regime and for clean substrates)
by the fact that the island density is not constant (as assumed in 
previous studies).

\section{Interpretation of experimental data}
\label{expdata}

In principle, Figure \ \ref{nmaxfnoscal} allows to determine the value 
of the microscopic parameters (diffusion, evaporation) if the saturation 
island density
is known. The problem is : does this island density correspond to the
defect concentration of the surface or to homogeneous nucleation? 
Is evaporation present in our experiments and what is the magnitude of $\tau_e$?

The first question can be answered by looking at the density evolution
with the flux. As already explained, if this
leaves unaffected the island density, nucleation is occuring on
defects. A similar test can be performed by changing the substrate 
temperature, but there is the nagging possibility that this changes the
defect concentration on the surface. It is also possible to study
the kinetics of island nucleation, i.e. look at the island
concentration as a function of thickness or coverage. The presence of
defects can be detected by the fact that the maximum island density is
reached at very low coverages (typically less than 1\%, see Fig. \
\ref{cinetns}).  One should be careful however to check that all the
islands, even those containing a few atoms, are visible in the
microscope images. 

The second question is more delicate. First, one should check 
whether atomic reevaporation is important. In principle, this can
be done by measuring the
sticking coefficient, i.e.  the amount of matter present on the
surface as a function of the matter brought by the beam. If possible, this
measure leaves no ambiguity. Otherwise, the kinetics of island
creation is helpful. If the saturation is reached at low thicknesses
($e_{sat} \leq .5 \ ML$), this means that evaporation is not
important. Another way of detecting atom evaporation is by studying the
evolution of the saturation island density with the
flux : the exponent goes from 0.29
to 0.66 (Fig. \ \ref{nmaxfnoscal}).  Suppose now that
one finds that evaporation is indeed important :
before being able to use Fig. \ \ref{nmaxfnoscal},
one has to know the precise value of $\tau_e$, and this is not
easily achieved.  For example, one could try to measure the
sticking coefficient or the quantity of matter needed to reach
saturation to obtain an estimation of the evaporation. Intuitively,
the more evaporation is important, the more matter we need to
reach the saturation density. Unfortunately, this strategy, although
useful for growth of 2d islands \cite{evaprb} is not so straightforward
here. The reason is that in the limit of strong evaporation (section \ \ref{strong}),
one has $e_{sat} \sim {N_{sat}}^{-1/2}$, thus bringing no independent 
information on the parameters. The same is true for the sticking
coefficient, which is a {\it constant}, i.e. independent of the value
of $\tau_e$ or the normalized flux. This counterintuitive result can be
understood by noting that in this limit, islands only grow by direct
impingement of atoms within them. Fortunately, in 
many experimental situations the limit of high evaporation is not reached 
and we "benefit" from
(mathematical) crossover regimes where these quantities do depend on
the precise values of $\tau_e$.  Fig. \ \ref{coll3d} gives the
evolutions of $S_{sat}$ and  $e_{sat}$ for
different values of $\tau_e$ and F in this crossover region. 
Then, a measurement of
$S_{sat}$ or $e_{sat}$ can shed light on experimental the value of $\tau_e$. 
For more details on interpretation of experimental data, we refer the reader
to a review paper to appear \cite{revmod}.

\section{Summary, Perspectives}

We have presented a comprehensive theoretical analysis of the growth of three 
dimensional structures on surfaces by atom evaporation. The study has been
carried out by combining a simple scaling analysis, a more rigorous
rate equations approach and computer simulations, with the main scope
of helping experimentalists to analyze their data. The scaling analysis can
give very simply the growth
exponents in the "clean" substrate case, in the limiting regimes of
high and low evaporation as well as in the crossover between these. The
rate equations confirm this
analysis and predict the growth on "dirty" substrates, i.e.
surfaces containing perfect traps for adatoms. The two approaches
were compared to Monte-Carlo computer simulations and very good
agreement was found. In addition to the analytical predictions, computer
simulations allowed to predict for the first time important growth
characteristics such as the island size distributions and intermediate
regimes which are difficult to study analytically. This is particularly
interesting for the interpretation of experimental data from the the 
sticking coefficient and the saturation island density (see section 
\ref{expdata}).

We wish to thank C. Henry for helpful discussions.

e-mail addresses : jensen@dpm.univ-lyon1.fr,
hernan@ce.ifisicam.unam.mx, Alberto.Pimpinelli@lasmea.univ-bpclermont.fr

\newpage

\begin{figure}
\caption{
{\it Typical island morphology generated by our model.}}
\label{pyramide}
\end{figure} 

\begin{figure}
\caption{ Universal function rescaling all our data. As predicted by
Equation \ \protect\ref{eqcross}, the normalized island density $N_{max}
X_S^2$ follows a single curve as a function of the evaporation
parameter $(F/D)^{2/3} {X_S}^{10/3} (1+ X_S^2)^{2/3}$.  The solid curve
shows the function predicted in the text (Equation \
\protect\ref{eqcross}), while the circles represent the results of
the computer simulations.}
\label{crossover}
\end{figure} 

\begin{figure}
\caption{
Evolution of the maximum island density as a function of the flux for
different evaporation times. The solid lines 
show the expected value for the exponent when evaporation is significant (2/3)
while the dashed line shows the exponent in the complete condensation case (2/7).}
\label{nmaxfnoscal}
\end{figure} 

\begin{figure}
\caption{
Maximum island density as a function of the evaporation time for different 
fluxes. The number next to each  symbol corresponds to the log(F) value 
for that set. The solid line shows the expected value for the exponent : 
2/3 for low values of $\tau_e$ (evaporation is significant).}
\label{nmaxtenoscal}
\end{figure}

\begin{figure}
\caption{
{\it Exponent of the mean island radius as a function of deposition
time in three deposition conditions : (a) complete condensation, (b)
strong evaporation (c) intermediate case. (a) shows the time exponent
for the radius (filled triangles) and for the islands (open squares).
We note that the exponent of the island evolution starts at 0.33 as predicted
but then slightly decreases to 0.26 (b) shows the averaged time exponent 
of the radius (open squares) as well
as the real mean values of the radius for 8 runs (to show the fluctuations).
The dashed line indicates the predicted value for the exponent (1).
(c) shows the values of the radius exponent for two different values
of the parameters, in the intermediate regime (high evaporation at the
beginning of the growth, decreasing as islands form, see below). There seems
to be a typical value for the exponent of about 0.6. In all these figures,
the value of the local exponent is obtained by a simple derivative in the 
log-log plot. The precise parameters used for each
graph are : (a) $F=10^{-11}, \tau_e=10^{15}, \tau=1 L=1500$, averaged over
5 runs; (b) $F=10^{-8}, \tau_e=1, \tau=1 L=1300$, averaged over 8 runs;
(c) squares : $F=10^{-10}, \tau_e=100, \tau=1 L=1550$, averaged over 3 runs,
filled circles : $F=10^{-8}, \tau_e=100, \tau=1 L=1350$, averaged over 8 runs}}
\label{rt}
\end{figure} 

\begin{figure}
\caption{Evolution of the island density as a function of the thickness
($e \equiv Ft$) for different growth hypothesis. This 
figure shows that the {\it same} saturation density can be obtained for
films grown in very different conditions. Note that the horizontal
scale is logarithmic : therefore, nucleation on defects
leads to saturation at extremely low coverages, almost impossible
to observe experimentally. The different sets of data represent :
{\it triangles} : growth with evaporation,
 $\tau_e = 100 \tau$ and $F\tau=1.2 \ 10^{-8}$, {\it circles} : growth
without evaporation ($F\tau=3 \ 10^{-10}$), and {\it squares} : 
growth on defects (defect concentration : $5 \ 10^{-4}$ per site) 
and $F\tau= 10^{-14}$ (no evaporation).}
\label{cinetns}
\end{figure} 

\begin{figure}
\caption{
Normalized island size distributions obtained for $F\tau=10^{-8}$ and
different values of the evaporation time $\tau_e$. The size distributions
are averaged for different coverages $\theta$ between .05 and 0.2. 
The solid line shows
the size distribution obtained without evaporation.  The number next
to each symbol corresponds to $\tau_e/\tau$.}
\label{distrib3d}
\end{figure} 

\begin{figure}
\caption{
Effect of the presence of defects on the island size distribution. The
rescaled island size distributions are obtained for $F\tau=10^{-8}$ and
different values of the evaporation time $\tau_e$ ($\tau =1$). The size
distributions were obtained for different coverages $\theta$ between
.05 and 0.15. Contrary to what is observed for homogeneous nucleation,
the histograms do depend on the coverage for nucleation on
defects. The solid line shows the size distribution obtained without
evaporation.}
\label{distribdef}
\end{figure}

\begin{figure}
\caption{
{\it Values of (a) the sticking coefficient $S_{sat}$ and (b) the thickness 
$e_{sat}$ at the {\it saturation} of island density in
 the total coalescence limit. In the limit of low island densities,
$S_{sat}$ is a {\it constant}.  However, there are crossover
regimes which depend on the precise $\tau_e$ and which are shown
here. Then, from a measure of $S_{sat}$ and $N_{sat}$ one can get an
estimate for $\tau_e$ for the not too low island densities which
correspond to many experimental cases. In the same spirit, (b) shows
the evolution of $e_{sat}$ as a function of $N_{sat}$ in the crossover
regime. The numbers correspond to the different $\tau_e/\tau$ used for
the simulations and the solid line represents the limiting regime 
(see \protect\cite{revmod} for more details).}}
\label{coll3d}
\end{figure}


\begin{thebibliography}{99}
\bibitem{software}
To help experimentalists, a software
simulating the submonolayer regime of thin-film growth in the various
growth hypothesis (2d or 3d islands, reversible aggregation, mobile
islands \ldots) is presently being developed and will soon be available 
by simple request to one of us (PJ).
\end{thebibliography}
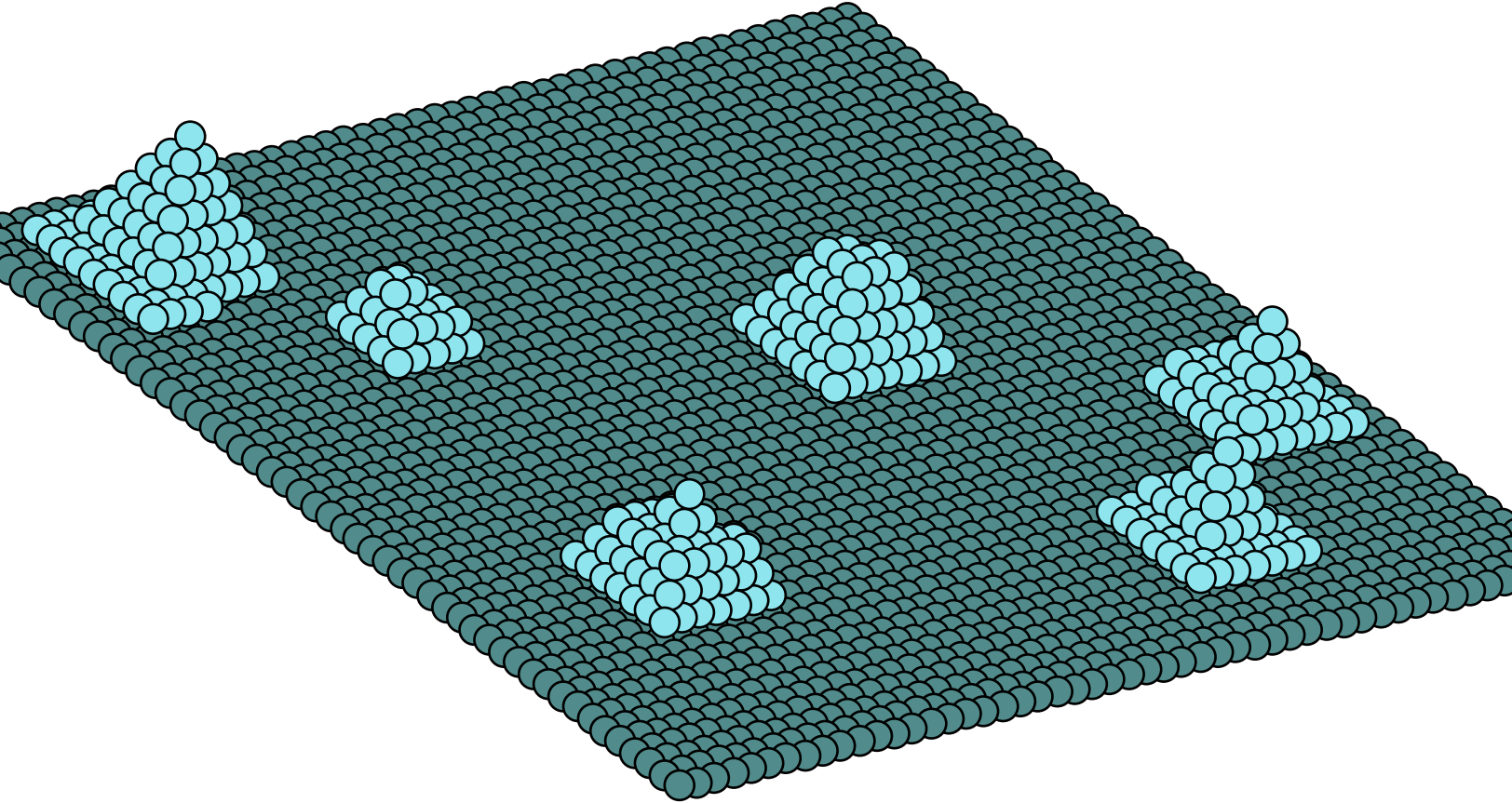
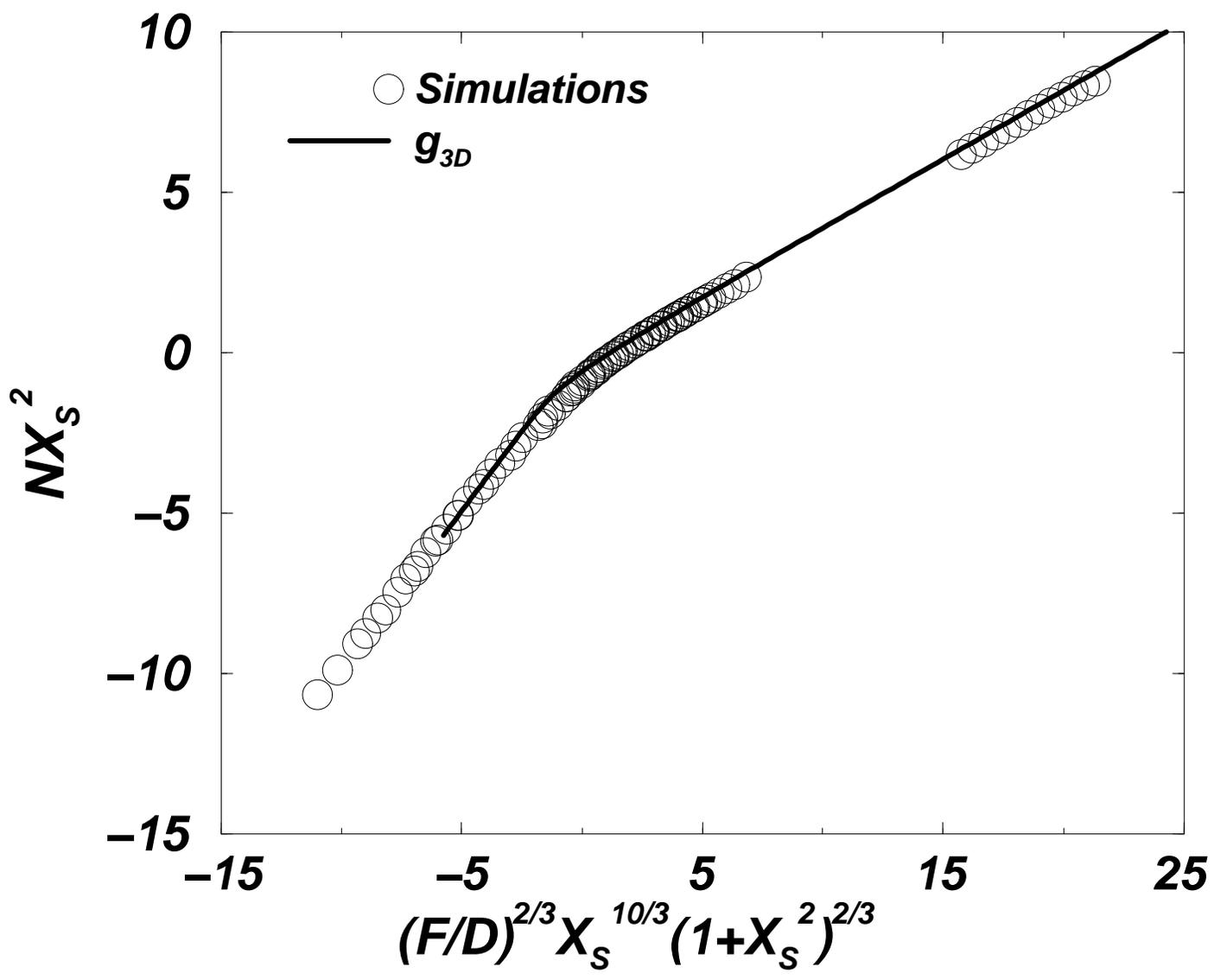
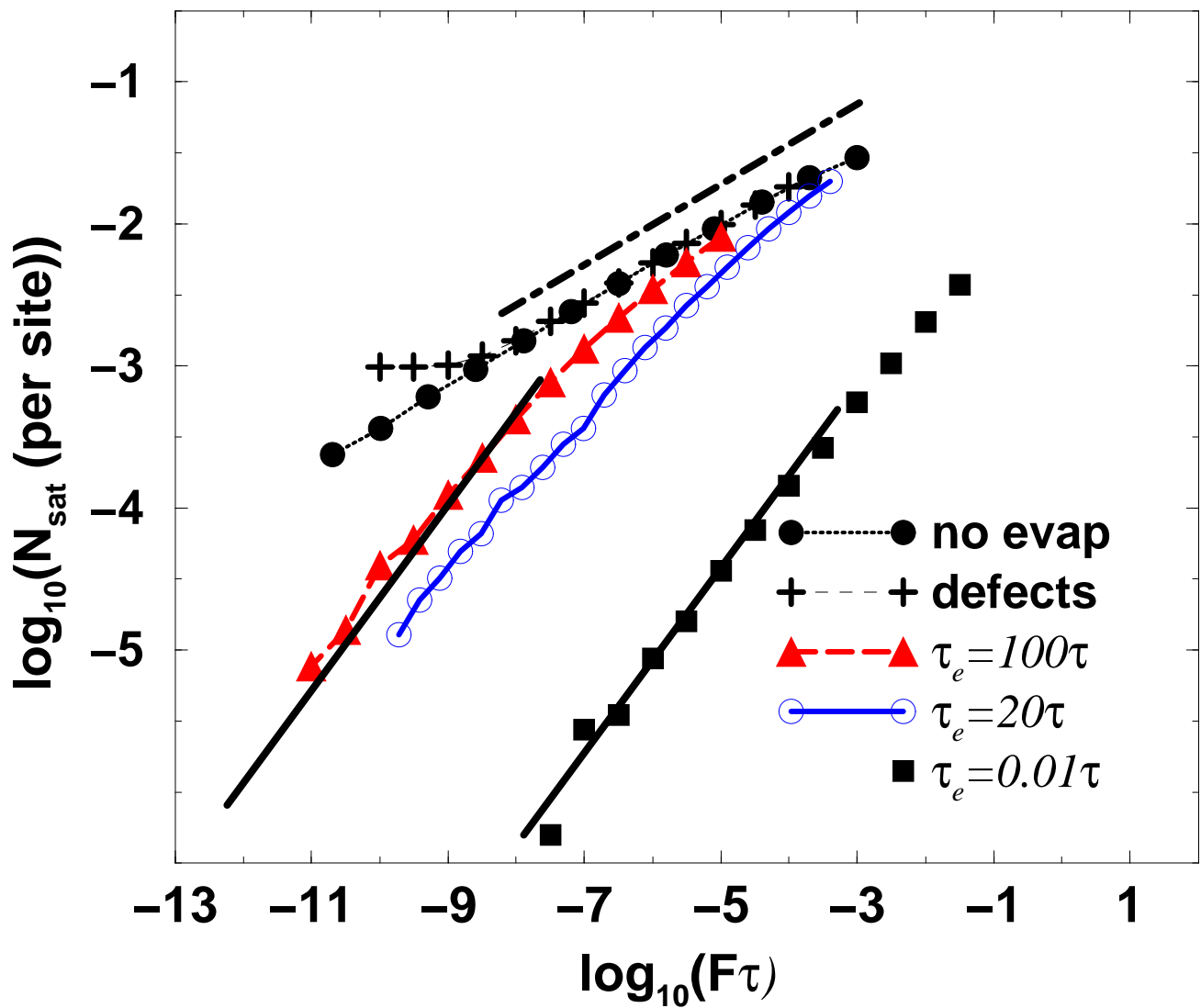
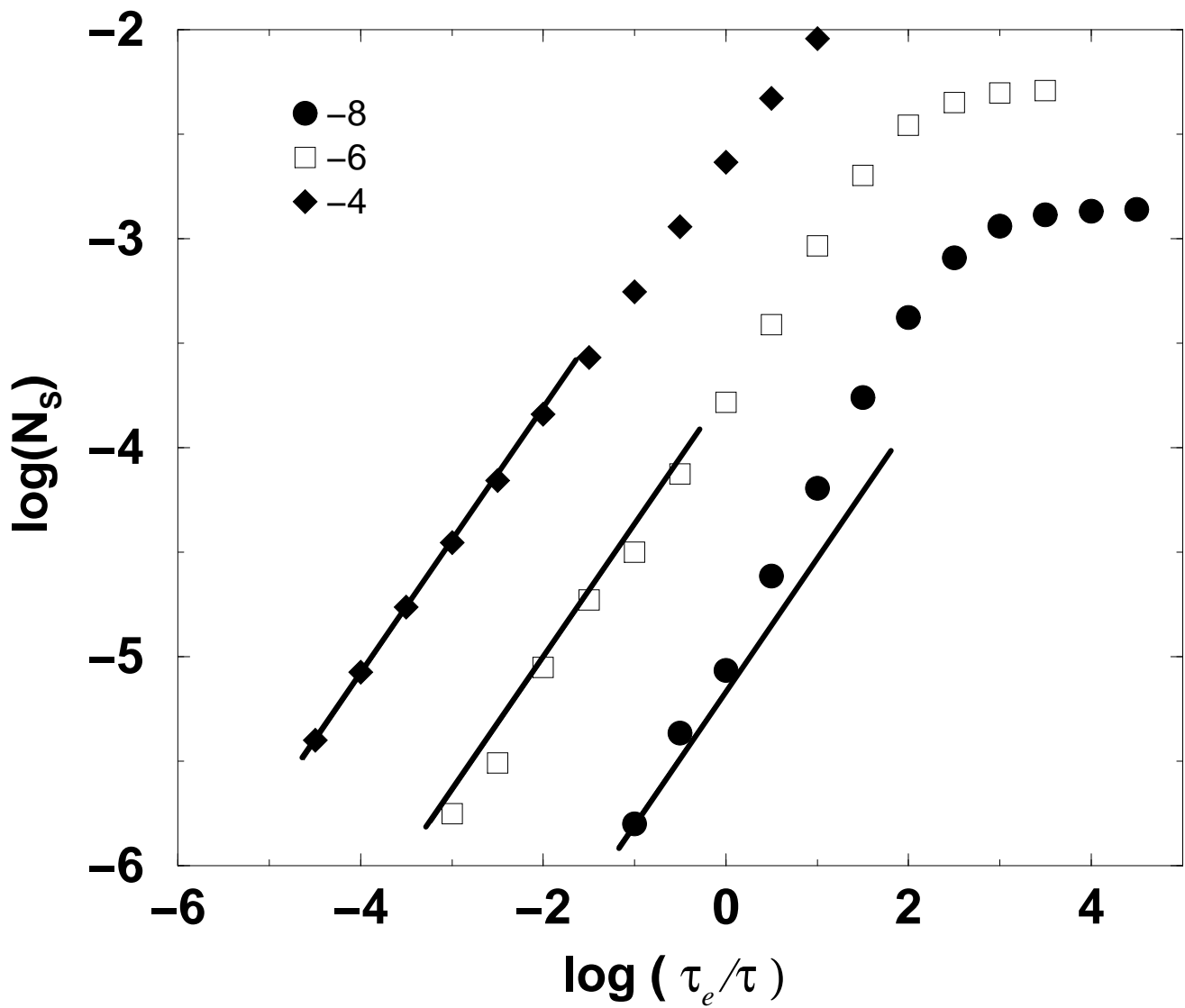
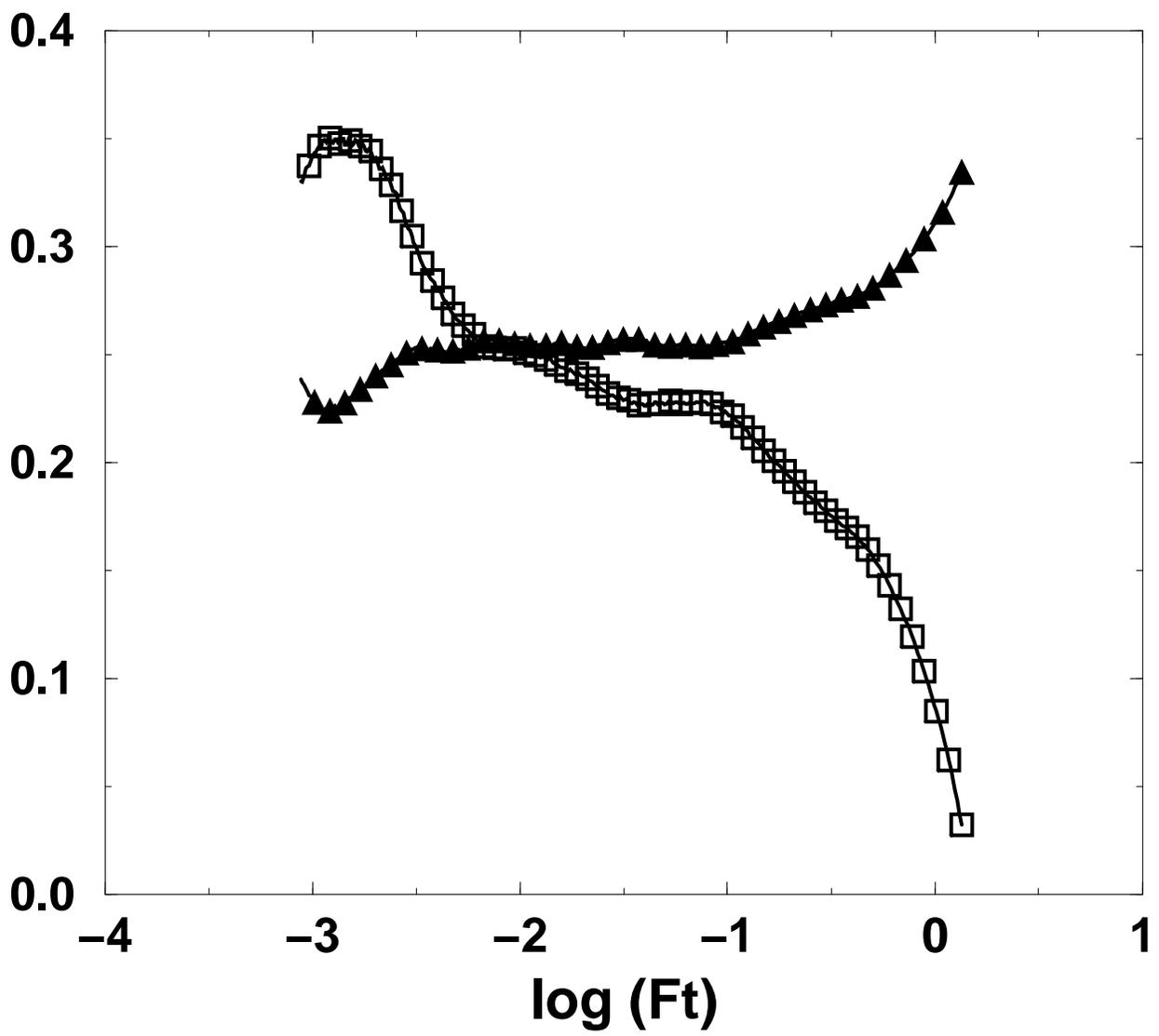
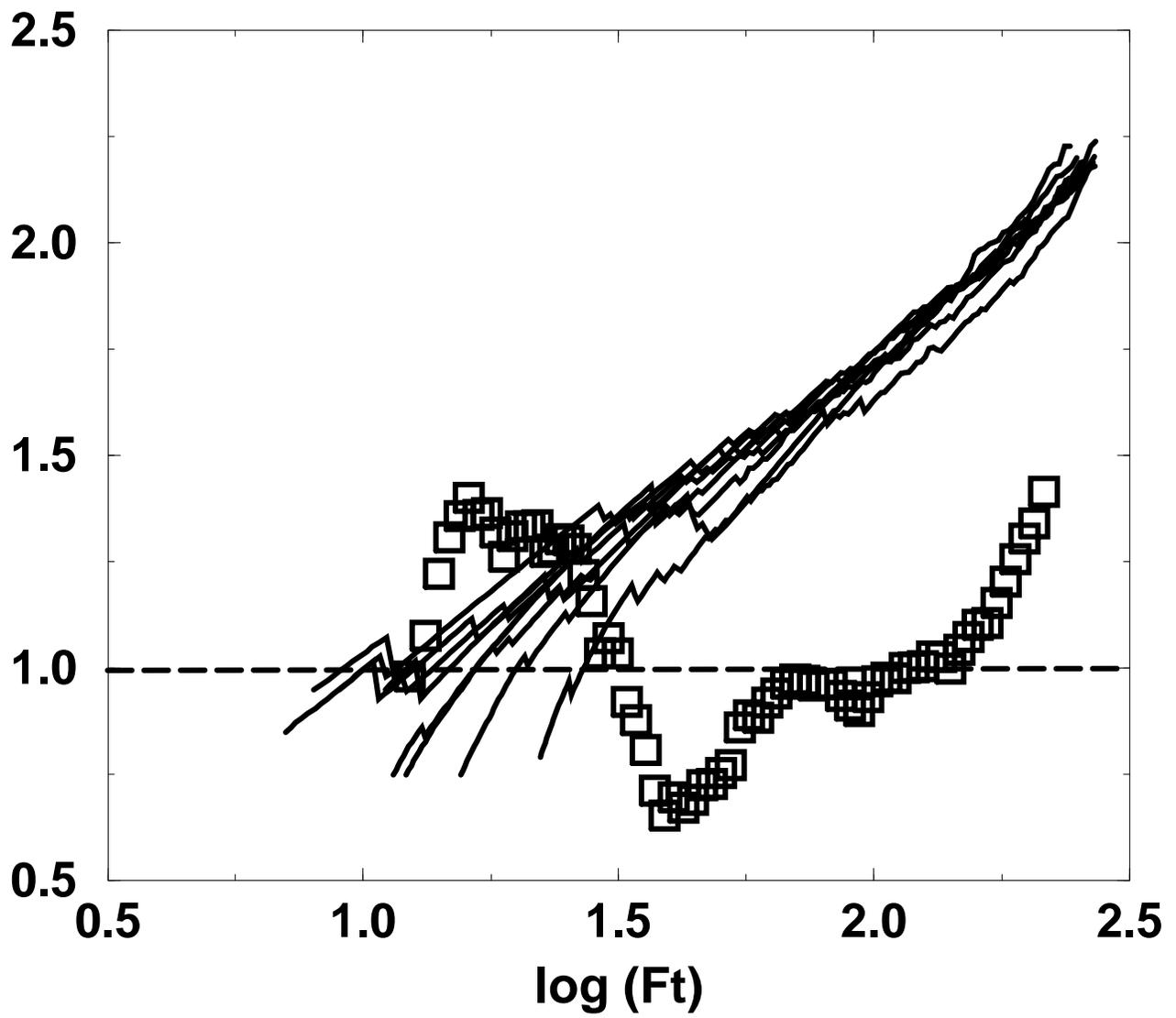
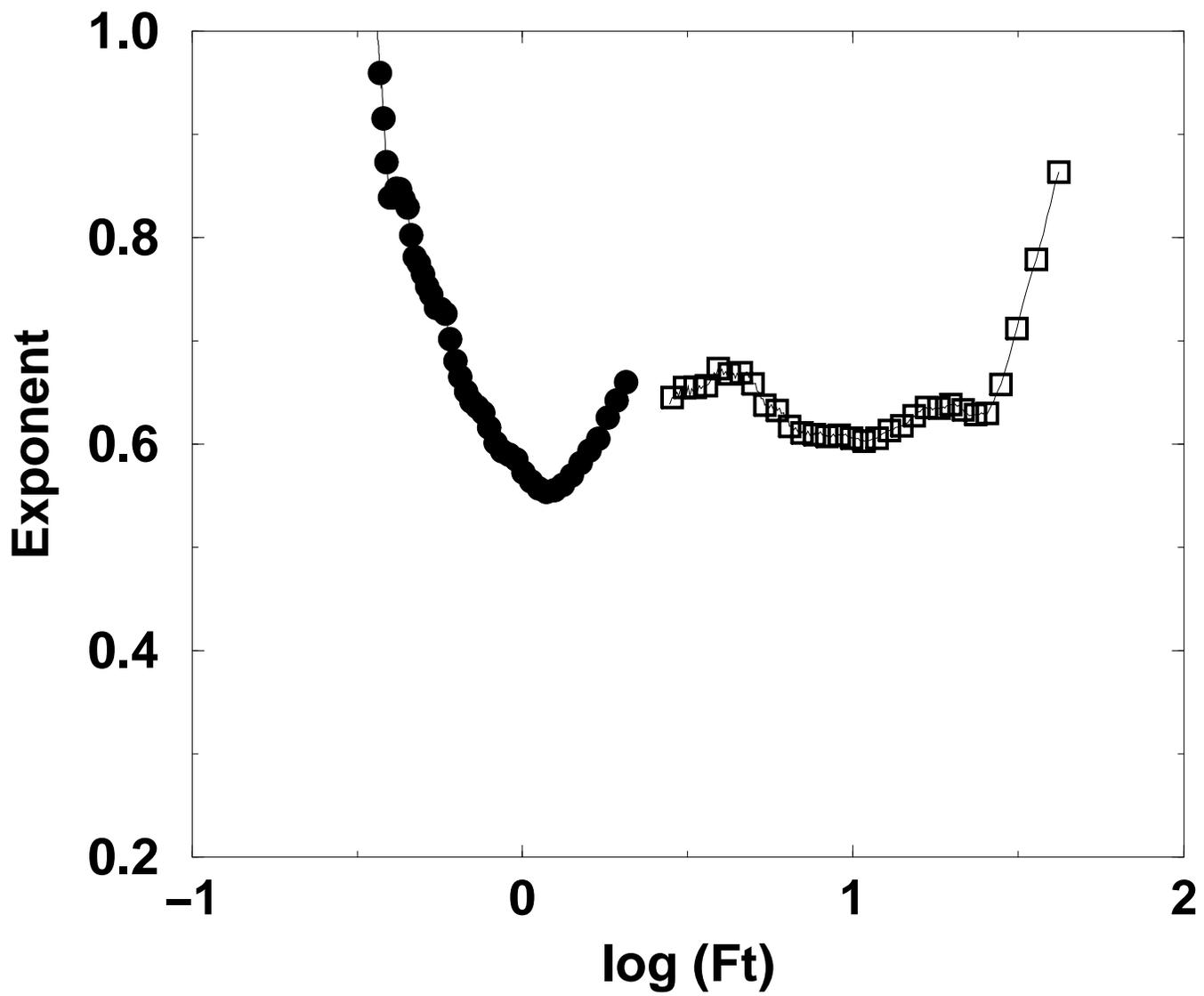
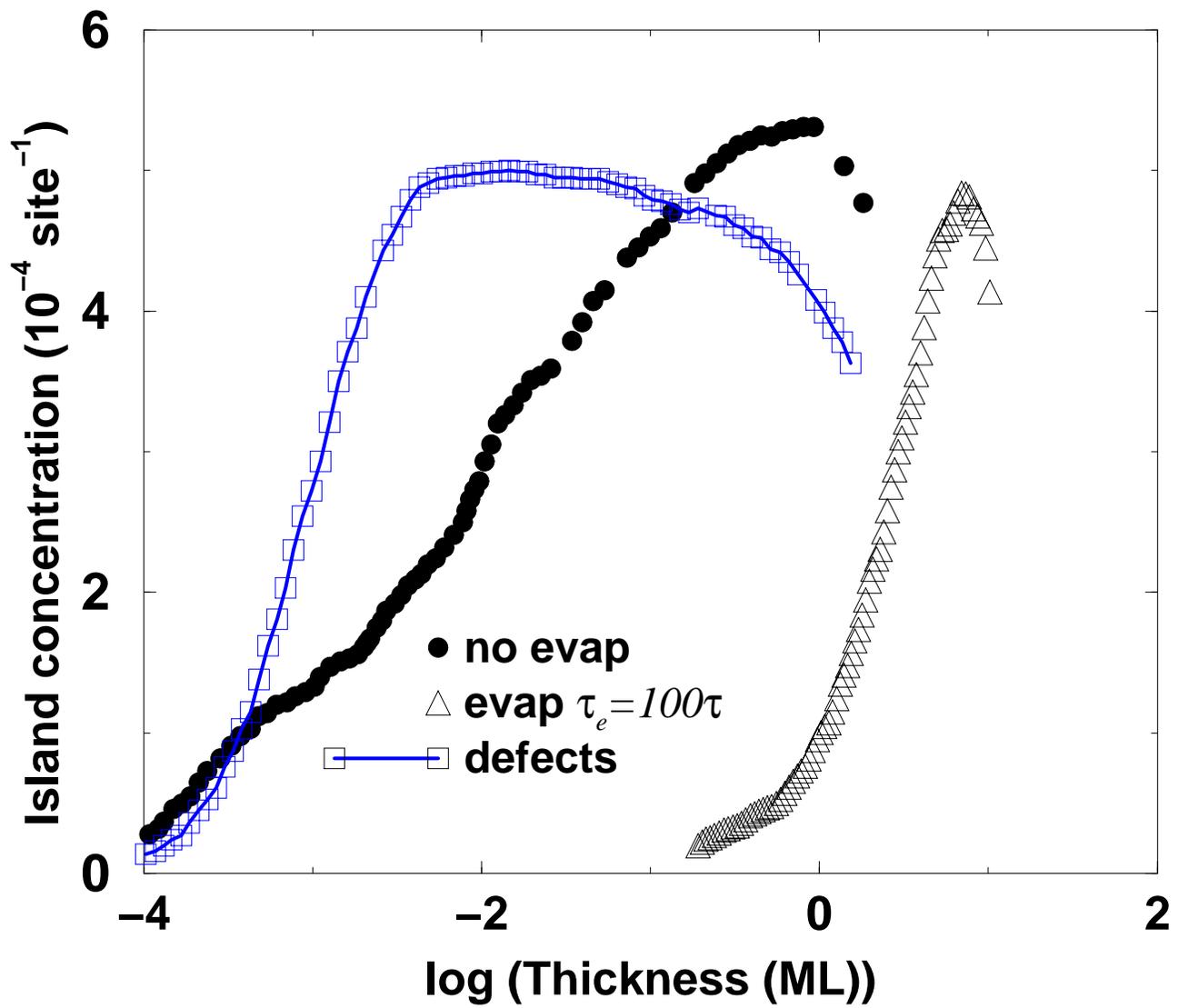
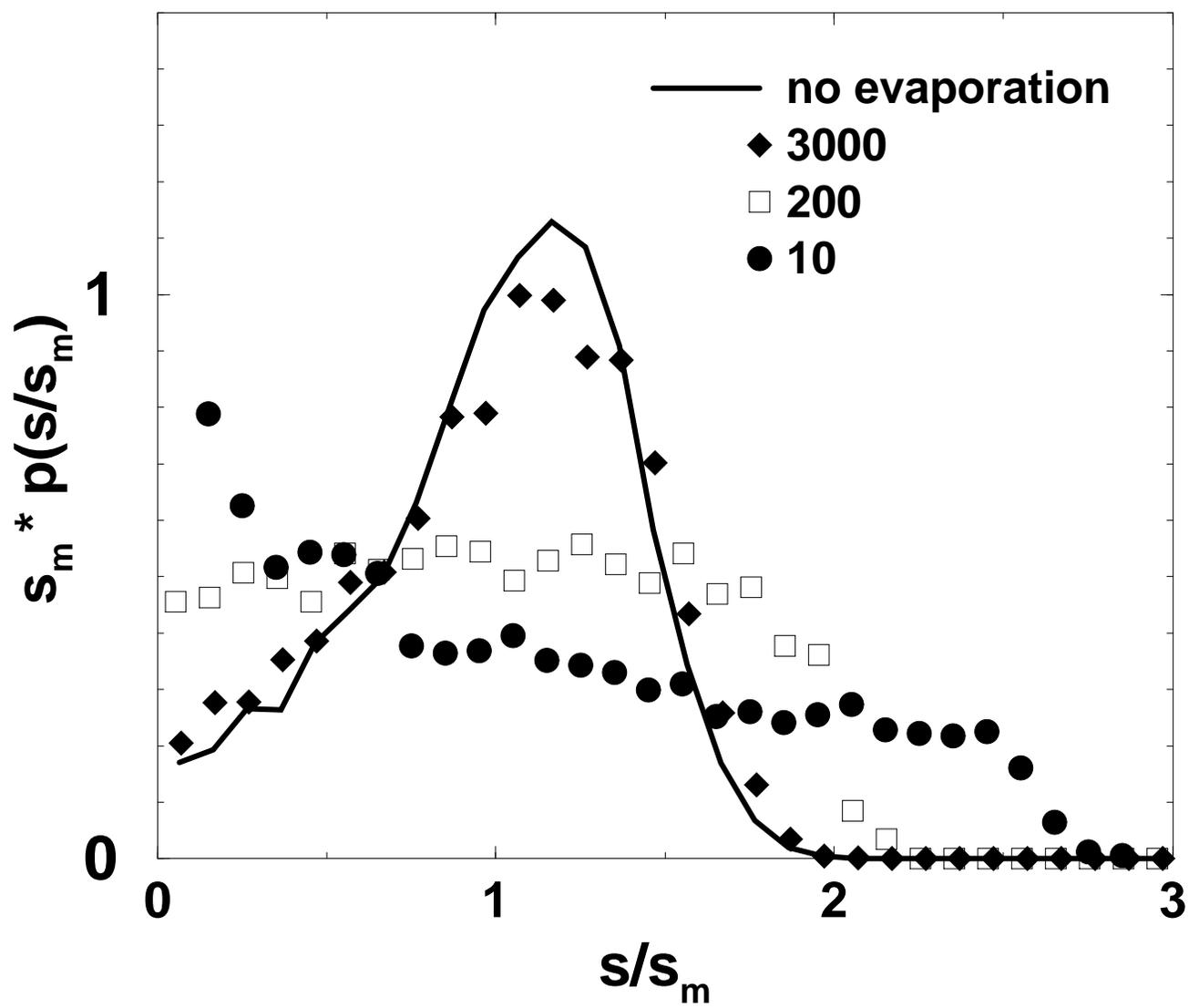
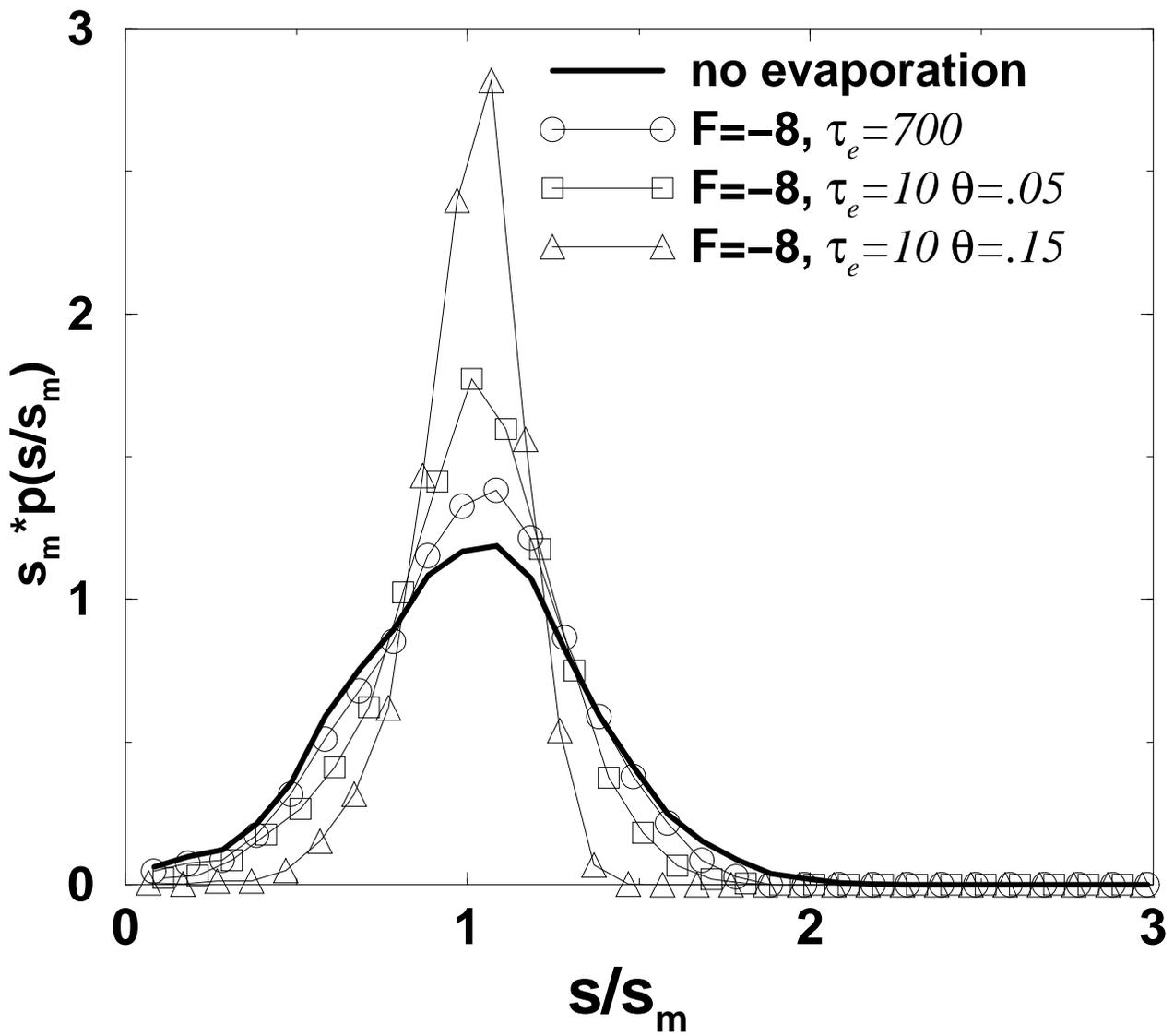
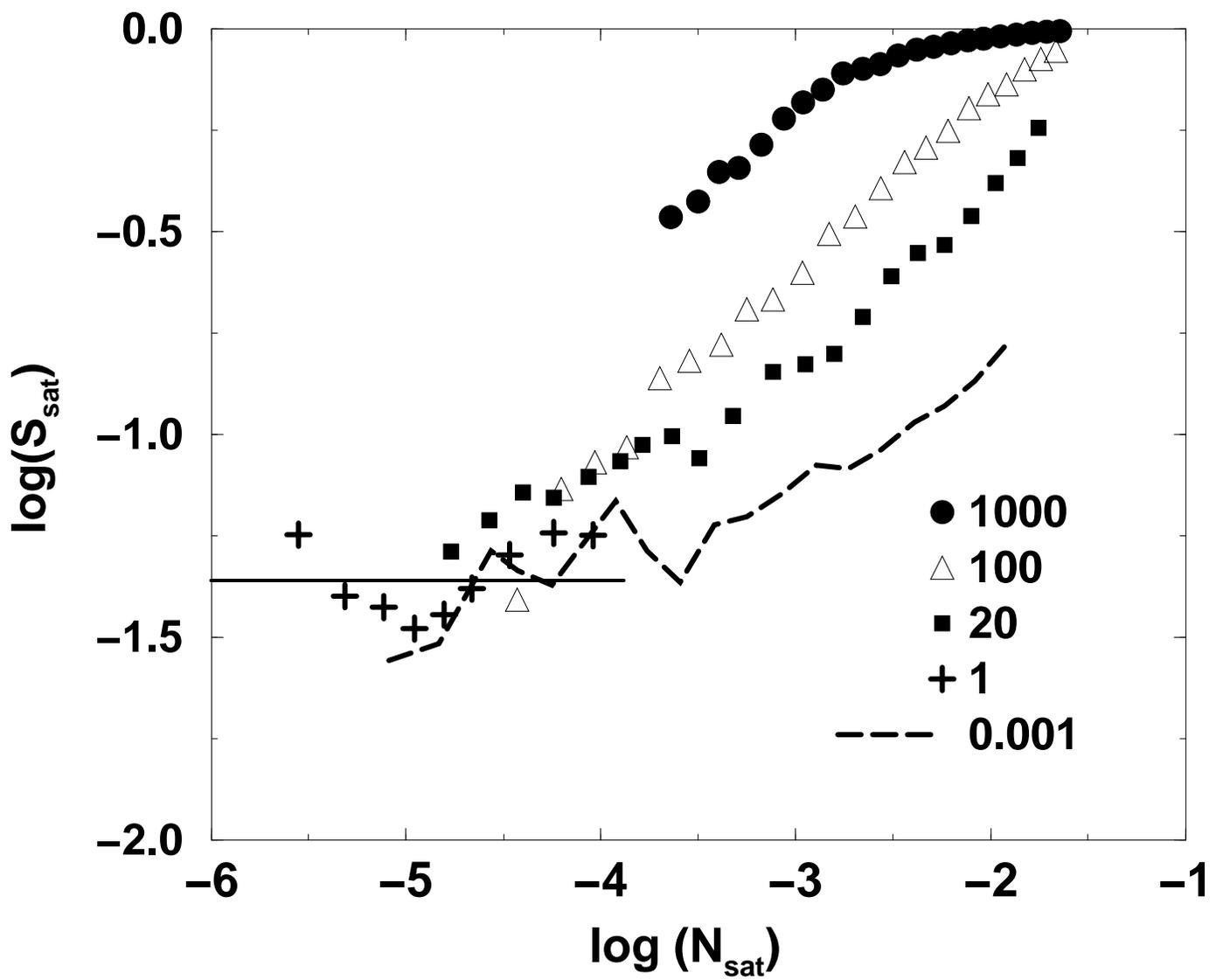
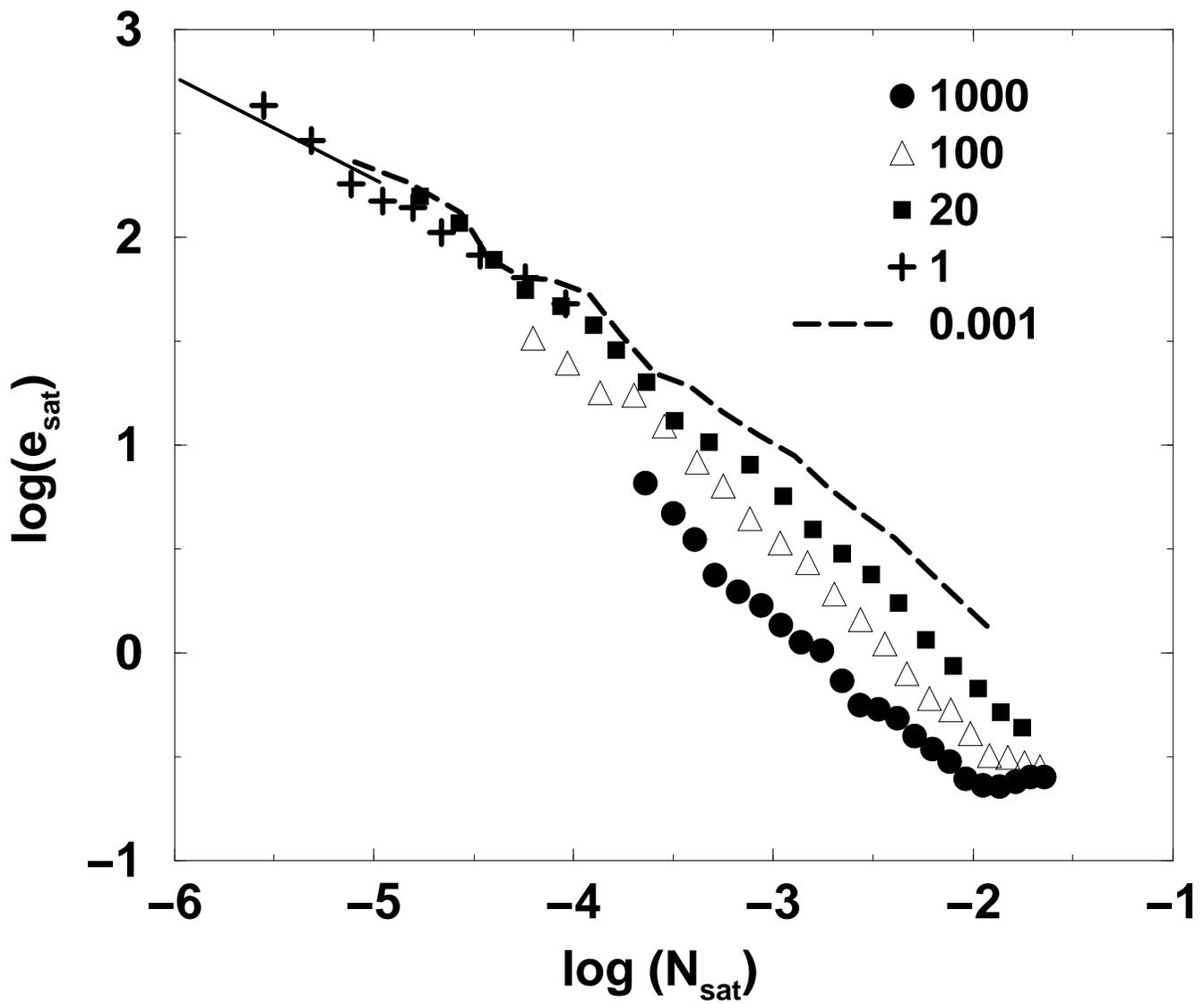
\end{document}